\documentclass[12pt,a4paper]{article}
\usepackage{amsmath,amssymb,graphicx,epsfig,cite,color}
\usepackage[left=2.7cm,right=2.7cm,top=2.5cm,bottom=2.5cm]{geometry}
\usepackage{subfigure}
\usepackage{multirow}
\usepackage[table]{xcolor}
\usepackage{booktabs}

\usepackage{xcolor,colortbl}

\usepackage{graphicx}
\usepackage{lscape}
\usepackage{appendix}
\usepackage{multirow}
\usepackage[colorlinks=true,linkcolor=blue,citecolor=blue,urlcolor=blue]{hyperref}%
\allowdisplaybreaks[1]

\begin{document}



\renewcommand{\textfraction}{0.2}    
\renewcommand{\topfraction}{0.8}   

\renewcommand{\bottomfraction}{0.4}   
\renewcommand{\floatpagefraction}{0.8}
\newcommand\mysection{\setcounter{equation}{0}\section}

\def\baeq{\begin{appeq}}     \def\eaeq{\end{appeq}}  
\def\baeeq{\begin{appeeq}}   \def\eaeeq{\end{appeeq}}
\newenvironment{appeq}{\beq}{\eeq}   
\newenvironment{appeeq}{\beeq}{\eeeq}
\def\bAPP#1#2{
 \markright{APPENDIX #1}
 \addcontentsline{toc}{section}{Appendix #1: #2}
 \medskip
 \medskip
 \begin{center}      {\bf\LARGE Appendix #1 :}{\quad\Large\bf #2}
\end{center}
 \renewcommand{\thesection}{#1.\arabic{section}}
\setcounter{equation}{0}
        \renewcommand{\thehran}{#1.\arabic{hran}}
\renewenvironment{appeq}
  {  \renewcommand{\theequation}{#1.\arabic{equation}}
     \beq
  }{\eeq}
\renewenvironment{appeeq}
  {  \renewcommand{\theequation}{#1.\arabic{equation}}
     \beeq
  }{\eeeq}
\nopagebreak \noindent}

\def\eAPP{\renewcommand{\thehran}{\thesection.\arabic{hran}}}

\renewcommand{\theequation}{\arabic{equation}}
\newcounter{hran}
\renewcommand{\thehran}{\thesection.\arabic{hran}}

\def\bmini{\setcounter{hran}{\value{equation}}
\refstepcounter{hran}\setcounter{equation}{0}
\renewcommand{\theequation}{\thehran\alph{equation}}\begin{eqnarray}}
\def\bminiG#1{\setcounter{hran}{\value{equation}}
\refstepcounter{hran}\setcounter{equation}{-1}
\renewcommand{\theequation}{\thehran\alph{equation}}
\refstepcounter{equation}\label{#1}\begin{eqnarray}}


\newskip\humongous \humongous=0pt plus 1000pt minus 1000pt
\def\caja{\mathsurround=0pt}
 

\title{
         {\Large
                 {\bf
Strong Vertices of Doubly Heavy Spin-3/2 Baryons with Light  Pseudoscalar Mesons  
                 }
         }
      }

\author{\vspace{1cm}\\
	{\small
		A. R. Olamaei$^{1,4}$,
		K. Azizi$^{2,,3,4}$,
		  S.~Rostami$^{2}$} \\
	{\small$^1$  Department of Physics, Jahrom University, Jahrom, P.~ O.~ Box 74137-66171, Jahrom, Iran}\\
	{\small $^2$ Department of Physics, University of Tehran, North Karegar Ave. Tehran 14395-547, Iran}\\
	{\small $^3$ Department of Physics, Do\v{g}u\c{s} University, Acibadem-Kadik\"{o}y, 34722
		Istanbul, Turkey}\\
	{\small $^4$ School of Particles and Accelerators, Institute for Research in Fundamental 
		Sciences (IPM),}\\
	{\small  P. O. Box 19395-5531, Tehran, Iran}\\
	} 

\date{}

\begin{titlepage}
\maketitle
\thispagestyle{empty}

\begin{abstract}
The strong coupling constants are basic quantities that carry information of the strong interactions among the baryon and meson multiplets as well as information on  the natures and internal structures  of the involved hadrons. These parameters enter to the transition matrix elements of various decays as main inputs and they play key roles in analyses of the experimental data including various hadrons. We determine the strong coupling constants among the doubly heavy spin-$ 3/2 $ baryons, $\Xi^*_{QQ'} $ and $\Omega^*_{QQ'}$, and light pseudoscalar mesons, $\pi$, $K$ and $\eta$, using the light-cone QCD. The values obtained for the strong coupling constants under study  may be used in construction of the strong potentials among the doubly heavy spin-3/2 baryons and light pseudoscalar mesons. 

\end{abstract}

\end{titlepage}
\section{Introduction}
Doubly heavy baryons composed of  two heavy quarks and 
a single light quark are interesting objects, 
 investigation of which can help us  better understand the  non-perturbative nature of QCD.
Their investigation  provides a good framework for understanding and predicting the spectrum
of heavy baryons. Theoretical studies on different aspects of these baryons may  shed light on the experimental searches of these states. These baryons have been in the focus of various theoretical studies \cite{Wang:2017mqp,Wang:2017azm,
Gutsche:2017hux,Li:2017pxa,Xiao:2017udy,Sharma:2017txj,Ma:2017nik,Hu:2017dzi,Shi:2017dto,Yao:2018ifh,Zhao:2018mrg,Liu:2018euh,Xing:2018lre,Dhir:2018twm,Berezhnoy:2018bde,Jiang:2018oak,Zhang:2018llc,Gutsche:2018msz,Shi:2019fph,Hu:2019bqj,Brodsky:2011zs,Mattson:2002vu,Ocherashvili:2004hi,Ratti:2003ez,Aubert:2006qw,Aaij:2013voa,Kato:2013ynr,Aliev:2012ru,Aliev:2012iv,Azizi:2014jxa,Aliev:2012nn,Ozdem:2019zis}. Some interesting progress on the OPE calculations of their lifetimes can be found in \cite{Onishchenko:1999yu,Chang:2007xa}.
Despite their predictions  decades ago via quark model, we have very limited experimental knowledge on these baryons. 
Nevertheless, the  year  2017 was special in this regard, because it was marked by the discovery of the doubly charmed $ \Xi^{++}_{cc} $ baryon
by the LHCb Collaboration \cite{Aaij:2017ueg}.  This state was then confirmed  in  the decay channel $\Xi^{++}_{cc}\rightarrow \Xi^+_c \pi^+$ \cite{Aaij:2018gfl}.   Although, the charmed-bottom state $ \Xi^0_{bc}$ has been searched in
the $\Xi^0_{bc}\rightarrow D^0 p K^- $ decay  by the LHCb collaboration,  no evidence for a signal is found \cite{Aaij:2020vid}. However, 
the detection of  $ \Xi^{++}_{cc} $ has raised hopes for the discovery of other members of doubly heavy baryons.
It is wonderful that the LHC opens new horizons in the discovery of heavy baryons  and
provides the possibility to investigate their electromagnetic, 
weak and strong decays.

Beyond the quark models \cite{GELLMANN1964214,Zweig:1981pd,Zweig:1964jf}, many  effective models like  lattice QCD \cite{Brown:2014ena}, Quark Spin Symmetry \cite{Hernandez:2007qv,Flynn:2011gf},  QCD Sum Rules \cite{Aliev:2012ru,Aliev:2012iv,Azizi:2014jxa,Aliev:2012nn,Ozdem:2019zis} and Light-Cone Sum Rules \cite{Shi:2019fph,Hu:2019bqj,Azizi:2018duk,Olamaei:2020bvw,Azizi:2020zin,Rostami:2020euc,Aliev:2020lly,Aliev:2020aon,Alrebdi:2020rev,Aliev:2021hqq} have been proposed to describe masses, residues, lifetimes, strong coupling constants, and other properties of doubly heavy baryons.  Given that the expectations to discover more doubly heavy baryons are growing, we are witnessing the rise of further theoretical investigations in this respect.

Concerning the strong decays of doubly heavy baryons, the  strong coupling constants are building blocks. They are entered  to the amplitudes of the strong decays and the widths of the related decays are calculated in terms of these constants. They can also be used to construct the strong  potential energy among the hadronic multiplets.
These couplings appear  in the low-energy (long-distance) region of QCD, where the running coupling gets larger and perturbation theory breaks down. Therefore, to calculate such coupling constants, a non-perturbative
approach should be used.
One  of the comprehensive and reliable methods for evaluating the non-perturbative effects is the light-cone QCD sum rule (LCSR) 
 which have given many successful descriptions of the hadronic properties so far. This method
is a developed version of the standard technique of SVZ  sum rule \cite{Shifman:1978bx},  using the conventional distribution amplitudes (DAs) of the on-sell states. The main difference between SVZ sum rule and LCSR is that the latter  employs the operator product expansion (OPE) near the light-cone  $ x^2\approx 0 $ instead of short distance $ x\approx 0 $, and the corresponding matrix elements are parametrized in terms of hadronic DAs which are classified according to their twists \cite{Balitsky:1989ry,Khodjamirian:1997ub,Braun:1997kw}.

In this study, we calculate the strong coupling constants among the doubly heavy spin-3/2 baryons, $\Xi^*_{QQ'}$ and $\Omega^*_{QQ'}$, and light pseudoscalar mesons $\pi$, $K$ and $\eta$. Here $ Q $ and $ Q' $ both can be
$ b $ or $ c $ quark. The paper is organized as follows: In the next section, we derive the sum rules for the strong coupling constants using the LCSR. Numerical analysis and results are presented in section \ref{NA}. The final section \ref{SC}, is devoted to the summary and conclusion.

\section{Light-Cone Sum Rules for $B^{*}_{1}B^{*}_{2}{\cal P}$ Strong Couplings }\label{LH}

In this section we aim to derive the sum rules for the strong coupling constants among the doubly heavy spin-3/2 baryons ($\Xi^{*}_{QQ'}$ and $\Omega^{*}_{QQ'}$) and light pseudoscalar mesons ($\pi$, $K$ and $\eta$) using the method of LCSR. 
The quark content of the doubly heavy spin-3/2 baryons are presented in Table \ref{tab:baryon3/2}.

\begin{table}
\setlength{\tabcolsep}{1.45em} 
\centering
\begin{tabular}{cccc}
\hline
\hline
\textcolor{blue}{Baryon}    &\textcolor{blue}{ $ q $ }& \textcolor{blue}{$ Q $ }& \textcolor{blue}{$ Q^\prime $ }\\
\hline
\hline
\\
$  \Xi^*_{QQ^\prime }$     & $ u $ or $ d $    & $b  $   or $ c $ & $b  $   or $ c $    \\
&         &  &     \\
$ \Omega^*_{QQ^\prime}$   & $ s $   & $b  $   or $ c $ & $b  $   or $ c $       \\
\\
\hline
\end{tabular}
\caption{The quark content of the doubly heavy spin-3/2 baryons.} \label{tab:baryon3/2} 
\end{table}

The first step to calculate the strong couplings is to write the proper correlation function (CF) in terms of doubly heavy baryons' interpolating current, $\eta_{\mu}$.  That is
\begin{eqnarray}\label{CF} 
\Pi_{\mu \nu}(p,q)= i \int d^4x e^{ipx} \left< {\cal P}(q) \vert {\cal T} \left\{
\eta_{\mu} (x) \bar{\eta}_{\nu} (0) \right\} \vert 0 \right>,
\end{eqnarray}
where ${\cal P}(q)$ represents the pseudoscalar meson carrying the four-momentum $q$, and $p$ represents the outgoing doubly heavy baryon four-momentum. The above CF can be calculated in two ways:
\begin{itemize}
	\item [$ \bullet $] One by inserting the complete set of hadronic states with the same quantum numbers of the corresponding doubly heavy baryons, which is called the physical or phenomenological side of the CF. It is calculated in the timelike region and contains observables like strong coupling constants.
	\item [$ \bullet $] In the second way, it  is calculated in the deep Euclidean spacelike region  by the help of OPE and in terms of DAs of the on-shell mesons and other QCD degrees of freedom. It is called the theoretical or QCD side of the CF.
\end{itemize}
 These two sides are matched via a dispersion integral which leads to the sum rules for the corresponding coupling constants. 
 To suppress the contributions of the higher states and continuum, Borel transformation and continuum subtraction procedures are applied.
 In the following, we explain each of these steps in detail. 

\subsection*{\textit{Constructing Interpolating Current}}
The procedure to construct the doubly heavy spin-3/2 interpolating current is to make a diquark structure having spin one and then attach the remaining spin-1/2 quark to that to build a spin-3/2 structure which has to contain the quantum numbers of the related baryon. Following the line of \cite{Aliev:2012iv} and \cite{Rostami:2020euc} one can find the corresponding interpolating current as follows.
The diquark structure has the form
\begin{eqnarray}
\eta_{\text{diquark}} = q_1^T C \Gamma q_2,
\end{eqnarray} 
where $T$ and $C$ represent the transposition and charge conjugation operators respectively and $\Gamma = I, \gamma_5, \gamma_{\mu}, \gamma_{\mu}\gamma_5, \sigma_{\mu\nu}$. The dependence of the spinors on the spacetime $x$ is omitted for now. Attaching the third quark, the general form of the corresponding structure is $[q_1^T C \Gamma q_2]\Gamma' q_3$. Considering the color indices ($a$,$b$ and $c$), the possible structures have the following forms: $\varepsilon_{abc}(Q^{aT} C \Gamma Q^{'b})\Gamma' q^c$, $\varepsilon_{abc}(q^{aT} C \Gamma Q^{b})\Gamma' Q^{'c}$ and $\varepsilon_{abc}(q^{aT} C \Gamma Q^{'b})\Gamma' Q^{c}$, where $Q^{(')}$ and $q$ are the heavy and light quark spinors respectively and the antisymmetric Levi-Civita tensor, $\varepsilon_{abc}$, makes the interpolating current color singlet.

The $\Gamma$ matrices can be determined by investigating the diquark part of the interpolating current. 
In the first form, as the diquark structure has spin 1, it must be symmetric under the exchange of the heavy quarks $Q \leftrightarrow Q'$. Transposing yields to 
\begin{equation}\label{curr1}
[\epsilon_{abc} Q^{aT} C\Gamma Q'^b ]^T =-\epsilon_{abc} Q'^{bT} \Gamma^T C^{-1}  Q^a= \epsilon_{abc} Q'^{bT} C(C\Gamma^T C^{-1}) Q^a,
\end{equation}
where we have used the identities $C^T = C^{-1}$ and $C^2 = -1$, and consider the anticommutation of the spinor components since they are Grassmann numbers. Since $C=i \gamma_2 \gamma_0$, the quantity $C\Gamma^T C^{-1}$ would be
\begin{equation}\label{curr2}
C\Gamma^T C^{-1} =
\begin{cases}
\Gamma & \text{for $\Gamma = 1, \gamma_5, \gamma_{\mu}\gamma_5 $,}\\
-\Gamma & \text{for $\Gamma = \gamma_{\mu}, \sigma_{\mu\nu}$.}
\end{cases}       
\end{equation}
Therefore, after switching the dummy indices in the LHS of Eq. (\ref{curr1}) one gets
\begin{equation}\label{curr3}
[\epsilon_{abc} Q^{aT} C\Gamma Q'^b ]^T =\pm \epsilon_{abc} Q'^{aT} C\Gamma  Q^b,
\end{equation}
where the $+$ and $-$ signs are for $\Gamma = \gamma_{\mu}, \sigma_{\mu \nu}$ and $ \Gamma=1, \gamma_5,  \gamma_5\gamma_\mu$ respectively. On the other hand, since the RHS of the above equation has to be symmetric with respect to the exchange of heavy quarks, we have
\begin{equation}\label{curr4}
[\epsilon_{abc} Q^{aT} C\Gamma Q'^b ]^T = \pm \epsilon_{abc} Q^{aT} C\Gamma  Q'^b,
\end{equation} 
where the $+$ and $-$ signs are for $\Gamma = \gamma_{\mu}, \sigma_{\mu \nu}$  and $ \Gamma=1, \gamma_5,  \gamma_5\gamma_\mu$ respectively.
Moreover, we know that the structure  $\epsilon_{abc} Q^{aT} C\Gamma Q'^b$ is a number ($1 \times 1$ matrix) and therefore the transposition leaves it unchanged. This leads us to conclude that the only possible choice for the $\Gamma$ matrices would be $\Gamma = \gamma_{\mu}$ or $\sigma_{\mu \nu}$.

The above mentioned symmetry of exchanging heavy quarks should be respected by the two remaining structures, $\varepsilon_{abc}(q^{aT} C \Gamma Q^{b})\Gamma' Q^{'c}$ and $\varepsilon_{abc}(q^{aT} C \Gamma Q^{'b})\Gamma' Q^{c}$, properly which leads to the following form:
\begin{eqnarray}\label{curr5}
\varepsilon_{abc}\big[(q^{aT} C \Gamma Q^{b})\Gamma' Q^{'c} + (q^{aT} C \Gamma Q^{'b})\Gamma' Q^{c}\big],
\end{eqnarray} 
where $\Gamma = \gamma_{\mu}$ or $\sigma_{\mu \nu}$.
Consequently, the interpolating current can be written in two possible forms as 
\begin{eqnarray}\label{curr6}
\varepsilon_{abc} \Big\{ \big(Q^{aT} C \gamma_{\mu} Q^{'b}\big)\Gamma'_1 q^c + \big(q^{aT} C \gamma_{\mu} Q^{b}\big)\Gamma'_1 Q^{'c} + \big(q^{aT} C \gamma_{\mu} Q^{'b}\big)\Gamma'_1 Q^{c} \Big\},
\end{eqnarray}
and
\begin{eqnarray}\label{curr7}
\varepsilon_{abc} \Big\{ \big(Q^{aT} C \sigma_{\mu \nu} Q^{'b}\big)\Gamma'_2 q^c + \big(q^{aT} C \sigma_{\mu \nu} Q^{b}\big)\Gamma'_2 Q^{'c} + \big(q^{aT} C \sigma_{\mu \nu} Q^{'b}\big)\Gamma'_2 Q^{c} \Big\}.
\end{eqnarray}
To determine $\Gamma'_1$ and $\Gamma'_2$, one should consider Lorentz and parity symmetries. As Eqs. (\ref{curr6}) and (\ref{curr7}) must have the Lorentz vector structure, $\Gamma'_1 = 1$ or $\gamma_5$ and $\Gamma'_2 = \gamma_{\nu}$ or $\gamma_{\nu}\gamma_5$. But parity considerations exclude the $\gamma_5$ matrix and therefore $\Gamma'_1 = 1$ and  $\Gamma'_2 = \gamma_{\nu}$. Moreover, Eq. (\ref{curr7}) won't survive if one consider all three quarks the same, and therefore the only possible choice comes from Eq. (\ref{curr6}) as
\begin{eqnarray}\label{curr8}
\eta_{\mu}(x) &=& \frac{1}{\sqrt{3}} \varepsilon_{abc} \Big\{ \big[Q^{aT}(x) C \gamma_{\mu} Q^{'b}(x)\big] q^c(x) + \big[q^{aT}(x) C \gamma_{\mu} Q^{b}(x)\big] Q^{'c}(x) \nonumber\\
&&+ \big[q^{aT}(x) C \gamma_{\mu} Q^{'b}(x)\big] Q^{c}(x) \Big\}.
\end{eqnarray}

Not that these currents couple not only to the positive-parity but to  negative-parity doubly-heavy baryons \cite{Azizi:2015ksa,Azizi:2015ica,Azizi:2015bxa,Aliev:2015qea,Aliev:2015xaa,Azizi:2015fqa}. Considering the  negative-parity partners bring  uncertainties in the numerical results  discussed also in \cite{Khodjamirian:2011jp}. Here, we do not consider the contribution of negative-parity baryons.

\subsection*{\textit{Physical Side}}
At hadronic (low energy) level, first we insert the complete set of hadronic states with the same quantum numbers of the corresponding initial and final doubly heavy spin-3/2 baryons and then perform the Fourier transformation by integrating over four-$x$. By isolating the ground state one gets
\begin{eqnarray}\label{hadron-level} 
\Pi^{\text{Phys.}}_{\mu\nu}(p,q)=\frac{\langle 0\vert \eta_\mu \vert B^*_2(p,r)\rangle  \langle B^*_2(p,r){\cal P}(q)\vert B^*_1(p+q,s)\rangle\langle B^*_1(p+q,s) \vert \bar{\eta}_\mu\vert 0\rangle}{(p^2-m_{2}^2)[(p+q)^2-m_{1}^2]} +\cdots~,
\end{eqnarray} 
where $B^*_1(p+q,s)$ and $B^*_2(p,r)$ are the incoming and outgoing doubly heavy spin-3/2 baryons with the masses $m_1$ and $m_2$ respectively. $s$ and $r$ are their spins and dots represent the higher states and continuum.
The matrix element $\langle 0\vert \eta_\mu\vert B^*_i(p,s)\rangle$ is defined as
\begin{eqnarray}\label{me1} 
\langle 0\vert \eta_\mu\vert B^*_i(p,s)\rangle &=&\lambda_{B^*_i}u_\mu(p,s),
\end{eqnarray}
where $u_\mu(p,s)$ is the Rarita--Schwinger spinor and $\lambda_{B^*_i}$ is the residue for the baryon $B^*_i$. The matrix element $\langle B^*_2(p,r){\cal P}(q)\vert B^*_1(p+q,s)\rangle$ can be determined using the Lorentz and parity considerations as 
\begin{eqnarray}\label{me2}
\langle B^*_2(p,r){\cal P}(q)\vert B^*_1(p+q,s)\rangle = g_{B^*_1 B^*_2 {\cal P}} \bar{u}_{\alpha}(p,r)\gamma_5 u^{\alpha}(p+q,s), 
\end{eqnarray}
where $g_{B^*_1 B^*_2 {\cal P}}$ is the strong coupling constant of the doubly heavy spin-3/2 baryons $B^*_1$ and $B^*_2$ with the light pseudoscalar meson ${\cal P}$.
After substituting the above matrix elements, (\ref{me1}) and (\ref{me2}), in (\ref{hadron-level}), since the initial and final baryons are unpolarized, one needs to sum over their spins using the following completeness relation:
\begin{eqnarray}
\label{SpinSum}
\sum_s u_\mu (p,s) \bar{u}_\nu (p,s) =- ( {\rlap/p +m } )\Bigg(
g_{\mu\nu} - {1\over 3} \gamma_\mu \gamma_\nu - {2 p_\mu p_\nu \over 3
	m^2} + {p_\mu \gamma_\nu - p_\nu \gamma_\mu \over 3 m} \Bigg)~,
\end{eqnarray} 
which may lead to the corresponding physical side of the CF.
But here we face two major problems: First, not all emerging Lorentz structures are independent; and second, since the interpolating current $\eta_{\mu}$ also couples to the spin-1/2 doubly heavy baryon states, there are some unwanted contributions from them which must be removed properly.
Imposing the condition $\gamma^{\mu}\eta_{\mu} = 0$, these contributions can be written as 
\begin{eqnarray}
\label{Spin1/2}
\langle 0 \vert \eta_{\mu}\vert B(p,s=1/2) \rangle = A \Big( \gamma_\mu - {4\over m_{\frac{1}{2}}}
p_\mu \Big) u(p,s=1/2)~.
\end{eqnarray}
To fix the above-mentioned problems, we re-order the Dirac matrices in a way that help us eliminate the spin-1/2 states' contributions easily. The ordering we choose is $\gamma_\mu \rlap/p \rlap/q \gamma_\nu\gamma_5$ which leads us to the final form of the physical side
\begin{eqnarray}
\label{CFPhys}
\Pi^{\text{Phys.}}_{\mu\nu}(p,q) &=& { \lambda_{B_{1}^*} \lambda_{B_{2}^*} \over
	[(p+q)^2-m_{1}^2)] (p^2 -  m_{2}^2)} \Big\{ \frac{2~ g_{B^*_1 B^*_2 {\cal P}}}{3 m_{1}^2} q_{\mu}q_{\nu}\rlap/p\rlap/q \gamma_5 +  \mbox{\rm  structures } \nonumber\\
	&&\mbox{beginning with $\gamma_\mu$ and ending with
	$\gamma_\nu\gamma_5$, or terms that are } \nonumber \\
& &\mbox{\rm proportional to $p_\mu$ or $(p+q)_\nu$}+\mbox{\rm other structures}\Big\} \nonumber\\
&&+ \int ds_1 ds_2 e^{-(s_1+s_2)/2M^2}\rho^{\text{Phys.}}(s_1,s_2),
\end{eqnarray}
where $\rho^{\text{Phys.}}(s_1,s_2)$ represents the spectral density for the higher states and continuum.
One should note that there are many structures like $g_{\mu\nu}\rlap/p\rlap/q \gamma_5$ that emerge in the QCD side of the CF which are not shown in the Eq. (\ref{CFPhys}) expansion and can be considered to calculate the strong coupling constant $g_{B^*_1 B^*_2 {\cal P}}$. We  select the structure with large number of momenta, $q_{\mu}q_{\nu}\rlap/p\rlap/q \gamma_5$,  which leads to more stable and reliable results  and it  is free of the unwanted doubly heavy spin-1/2 contributions.

Now, to suppress the contributions of the higher states and continuum, we perform the double Borel transformation with respect to the squared momenta $p_1^2 = (p+q)^2$ and $p_2^2 = p^2$, which leads to
\begin{eqnarray}\label{CFPhysB}
{\cal B}_{p_1}(M_1^2){\cal B}_{p_2}(M_2^2)\Pi^{\text{Phys.}}_{\mu\nu}(p,q) &\equiv & \Pi^{\text{Phys.}}_{\mu\nu}(M^2)\\
&=& \frac{2~ g_{B^*_1 B^*_2 {\cal P}}}{3 m_{1}^2} \lambda_{B_1} \lambda_{B_2} e^{-m_{1}^2/M_1^2} e^{-m_{2}^2/M_2^2}q_{\mu}q_{\nu}\rlap/p\rlap/q \gamma_5~ + \cdots~, \nonumber
\end{eqnarray}
where dots represent suppressed  higher states and continuum contributions, $M_1^2$ and $M_2^2$ are the Borel parameters and $ M^2= M^2_1 M^2_2/(M^2_1+M^2_2) $. The Borel parameters are chosen to be equal since the mass of the initial and final baryons are the same, therefore $ M^2_1 = M^2_2 = 2M^2 $. After matching the physical and QCD sides, we will perform the continuum subtraction supplied by quark-hadron duality assumption.

\subsection*{\textit{QCD Side}}

After evaluating the physical side of the CF, the next stage is to calculate the QCD side in the deep Euclidean spacelike region,  where $-(p+q)^2\rightarrow \infty$ and $-p^2\rightarrow \infty$. The CF in this way is calculated in terms of QCD degrees of freedom as well as  non-local matrix elements  of pseudoscalar mesons expressed in terms of the DAs of different twists  \cite{Ball:2005vx,Ball:2004ye,Ball:1998je}.

To proceed, according to Eq. (\ref{CFPhys}), we choose the relevant structure $q_{\mu}q_{\nu}\rlap/p\rlap/q \gamma_5$ from the QCD side  and express the CF as
\begin{eqnarray}\label{QCD}
\Pi^{\text{QCD}}_{\mu\nu}(p,q) = \Pi(p,q) q_{\mu}q_{\nu}\rlap/p\rlap/q \gamma_5,
\end{eqnarray}
where $\Pi(p,q)$ is an invariant function of $(p+q)^2$ and $p^2$.
To calculate $\Pi(p,q)$, we insert the interpolating current (\ref{curr8}) into the CF (\ref{CF}), and using the Wick theorem we find the QCD side of the CF as:
\begin{eqnarray}\label{CF1}
\Big(\Pi^{\text{QCD}}_{\mu \nu}\Big)_{\rho\sigma}(p,q) &=&  \frac{i}{3}\epsilon_{abc}\epsilon_{a'b'c'} \int d^4 x e^{i q.x} \langle {\cal P}(q) \vert \bar{q}^{c^\prime}_{\alpha}(0)q^{c}_{\beta}(x)\vert 0\rangle \nonumber\\
&&\times \Bigg\{\delta_{\rho \alpha} \delta_{\beta \sigma }\text{Tr} \Big[  \tilde{S}^{aa^{\prime}}_{Q}(x) \gamma_\mu S^{bb^{\prime}}_{Q^{\prime}}(x) \gamma_\nu  \Big] +
\delta_{\alpha \rho}\Big(  \gamma_\nu \tilde{S}^{aa^{\prime}}_{Q}(x) \gamma_\mu S^{bb^{\prime}}_{Q^{\prime}}(x) \Big)_{\beta \sigma } \nonumber\\
&&+
\delta_{\alpha \rho} \Big( \gamma_\nu   \tilde{S}^{bb^{\prime}}_{Q^{\prime}}(x) \gamma_\mu S^{aa^{\prime}}_{Q}(x) \Big)_{\beta \sigma } + \delta_{\beta\sigma} \Big(  S^{bb^{\prime}}_{Q^{\prime}}(x)  \gamma_\nu \tilde{S}^{aa^{\prime}}_{Q}(x)  \gamma_\mu \Big)_{ \rho \alpha } \nonumber\\
&&-\delta_{\beta \sigma}\Big( S^{aa^{\prime}}_{Q}(x)  \gamma_\nu   \tilde{S}^{bb^{\prime}}_{Q^{\prime}}(x) \gamma_\mu  \Big)_{ \rho \alpha} +
\Big(  \gamma_\nu \tilde{S}^{aa^{\prime}}_{Q}(x)  \gamma_\mu   \Big)_{ \beta \alpha} \Big(S^{bb^{\prime}}_{Q^{\prime}}(x)\Big)_{\rho \sigma} \nonumber\\ &&- \Big( C  \gamma_\mu S^{aa^{\prime}}_{Q}(x)  \Big)_{\alpha \sigma} \Big(   S^{bb^{\prime}}_{Q^{\prime}}(x) \gamma_\nu C \Big)_{\rho \beta} -\Big( C  \gamma_\mu S^{bb^{\prime}}_{Q^{\prime}}(x)  \Big)_{\alpha \sigma}  \Big(   S^{aa^{\prime}}_{Q}(x) \gamma_\nu C \Big)_{\rho \beta} \nonumber\\
&&+
\Big(\gamma_\nu  \tilde{S}^{bb^{\prime}}_{Q^{\prime}}(x)  \gamma_\mu \Big)_{\beta \alpha } \Big(S^{aa^{\prime}}_{Q}(x)\Big)_{\rho \sigma}\Bigg\},
\end{eqnarray}
where $S^{aa'}_{Q}(x)$ is the propagator of the heavy quark $Q$ with color indices $a$ and $a'$, and $\tilde{S} = C S^T C$. $\mu$ and $\nu$ are Minkowski and $\rho$ and $\sigma$ are Dirac indices respectively. The propagators in the above equation contain both the perturbative and the non-perturbative contributions. As we previously mentioned, the  non-local matrix elements $\langle{\cal P}(q)\vert \bar{q}^{c^\prime}_{\alpha}(0)q^{c}_{\beta}(x)\vert 0\rangle$, are written in terms of DAs of the corresponding light pseudoscalar meson ${\cal P}(q)$.

The next stage is to insert the explicit expression of the heavy quark propagator in Eq. (\ref{CF1})  as \cite{Aliev:2011ufa}:
\begin{eqnarray}\label{HQP}
S_Q^{aa^{\prime}}(x) &=& {m_Q^2 \over 4 \pi^2} {K_1(m_Q\sqrt{-x^2}) \over \sqrt{-x^2}}\delta^{aa^{\prime}} -
i {m_Q^2 \rlap/{x} \over 4 \pi^2 x^2} K_2(m_Q\sqrt{-x^2})\delta^{aa^{\prime}}\nonumber \\ &&-
ig_s \int {d^4k \over (2\pi)^4} e^{-ikx} \int_0^1
du \Bigg[ {\rlap/k+m_Q \over 2 (m_Q^2-k^2)^2} \sigma^{\lambda\tau} G_{\lambda\tau}^{aa^{\prime}} (ux)
\nonumber \\
&&+
{u \over m_Q^2-k^2} x^\lambda  \gamma^\tau G_{\lambda\tau}^{aa^{\prime}}(ux) \Bigg]+\cdots,
\end{eqnarray} 
where $m_Q$ is the heavy quark mass, $K_1$ and $K_2$ are the modified Bessel functions of the second kind and  $G_{\lambda\tau}^{aa^{\prime}}$ is the gluon field strength tensor. It is defined as
\begin{eqnarray}
G^{aa^{\prime}}_{\lambda\tau }\equiv G^{A}_{\lambda\tau}t^{aa^{\prime}}_{A},
\end{eqnarray}
with $\lambda$ and $\tau$ being the Minkowski indices. $t^{aa'}_{A} = \lambda^{aa'}_{A}/2$ where $\lambda_{A}$ are the Gell-Mann matrices with $A=1, \cdots, 8$ and $a, a'$ the color indices. 
The free propagator contribution is determined by the first two terms and the rest which $\sim G_{\lambda\tau}^{aa^{\prime}}$ are due to the interaction with the gluon field.

Inserting Eq. (\ref{HQP}) into (\ref{CF1}) leads to several contributions. The first one is due to replacing both the heavy quark propagators with their free part:
\begin{eqnarray}\label{HQPpert}
S_Q^{(\text{pert.})}(x) &=& {m_Q^2 \over 4 \pi^2} {K_1(m_Q\sqrt{-x^2}) \over \sqrt{-x^2}} -
i {m_Q^2 \rlap/{x} \over 4 \pi^2 x^2} K_2(m_Q\sqrt{-x^2}).
\end{eqnarray} 
 The DAS in this case are two-particle DAs. Replacing one heavy quark propagator (say $S^{aa'}_{Q}$) with its gluonic part
\begin{eqnarray}\label{HQPnp}
S^{aa^{\prime}(\text{non-p.})}_{Q}(x)&=&-
ig_s \int {d^4k \over (2\pi)^4} e^{-ikx} \int_0^1
du G^{aa^{\prime}}_{\lambda\tau}(ux) \Delta^{\lambda\tau}_{Q}(x),
\end{eqnarray}
where
\begin{eqnarray}\label{HQPgamma}
\Delta^{\lambda\tau}_{Q}(x)&=& \dfrac{1}{2 (m_Q^2-k^2)^2}\Big[(\rlap/k+m_Q)\sigma^{\lambda\tau} + 2u (m_Q^2-k^2)x^\lambda \gamma^\tau\Big],
\end{eqnarray}
and the other with its free part, leads to the one gluon exchange between the heavy quark $Q$ and the light pseudoscalar meson ${\cal P}$. The non-local matrix elements of this contribution can be calculated in terms of three-particle DAs of meson $\cal P$. Replacing both heavy quark propagators with their gluonic parts involves ${\cal P}$ meson four-particle DAs which are not yet determined and we ignore them in the present work. Instead, we consider the two-gluon condensate contributions in this study. 

The non-local matrix elements can be expanded using proper Fierz identities like
\begin{eqnarray}\label{eq:Fierz1}
\bar q _\alpha ^{c'} q _\beta ^c  \to - \frac{1}{12} (\Gamma _J) _{\beta\alpha} \delta ^{cc'} \bar q \Gamma^J q , 
\end{eqnarray}
where $\Gamma ^{J}=\mathbf{1,\ }\gamma _{5},\ \gamma _{\mu },\ i\gamma _{5}\gamma
_{\mu },\ \sigma _{\mu \nu }/\sqrt{2}$. It puts them in a form to be determined  in terms of the corresponding DAs of different twists which can be found in Refs. \cite{Ball:2005vx,Ball:2004ye,Ball:1998je}.

Putting all things together, one can calculate different contributions to the corresponding strong coupling. The leading order contribution corresponding to no gluon exchange which can be obtained by replacing both heavy quark propagators with their free part is as follows:
\begin{eqnarray}\label{CFpert}
\Big(\Pi^{\text{QCD(0)}}_{\mu \nu}\Big)_{\rho\sigma}(p,q) &=&  \frac{i}{6} \int d^4 x e^{i q.x} \langle {\cal P}(q) \vert \bar{q}(0) \Gamma^J q(x)\vert 0\rangle \Bigg\{\text{Tr}\Big[  \tilde{S}^{(\text{pert.})}_{Q}(x) \gamma_\mu S^{(\text{pert.})}_{Q^{\prime}}(x) \gamma_\nu  \Big]\Big(\Gamma_{J}\Big)_{\rho \sigma} \nonumber\\ &+&
\text{Tr}  \Big[ \Gamma_J \gamma_\nu \tilde{S}^{(\text{pert.})}_{Q}(x)  \gamma_\mu   \Big] \Big(S^{(\text{pert.})}_{Q^{\prime}}(x)\Big)_{\rho \sigma} +
\text{Tr} \Big[\Gamma_J \gamma_\nu  \tilde{S}^{(\text{pert.})}_{Q^{\prime}}(x)  \gamma_\mu \Big] \Big(S^{(\text{pert.})}_{Q}(x)\Big)_{\rho \sigma} \nonumber\\ &+&
\Big(\Gamma_J  \gamma_\nu \tilde{S}^{(\text{pert.})}_{Q}(x) \gamma_\mu S^{(\text{pert.})}_{Q^{\prime}}(x) \Big)_{\rho \sigma } +
\Big( \Gamma_J \gamma_\nu   \tilde{S}^{(\text{pert.})}_{Q^{\prime}}(x) \gamma_\mu S^{(\text{pert.})}_{Q}(x)  \Big)_{\rho \sigma } 
\nonumber\\
&+&
\Big(  S^{(\text{pert.})}_{Q^{\prime}}(x)  \gamma_\nu \tilde{S}^{(\text{pert.})}_{Q}(x)  \gamma_\mu \Gamma_J\Big)_{ \rho \sigma } 
-\Big( S^{(\text{pert.})}_{Q^{\prime}}(x) \gamma_\nu  \tilde{\Gamma}_J  \gamma_\mu S^{(\text{pert.})}_{Q}(x)  \Big)_{\rho \sigma} \nonumber\\ 
&-&\Big( S^{(\text{pert.})}_{Q}(x)  \gamma_\nu   \tilde{S}^{(\text{pert.})}_{Q^{\prime}}(x) \gamma_\mu  \Gamma_J \Big)_{ \rho \sigma}-\Big( S^{(\text{pert.})}_{Q}(x) \gamma_\nu \tilde{\Gamma}_J  \gamma_\mu S^{(\text{pert.})}_{Q^{\prime}}(x)  \Big)_{\rho \sigma}  
\Bigg\},
\end{eqnarray}
where the superscript $(0)$ indicates no gluon exchange. The contribution of the exchange of one gluon between the heavy quark $Q$ and the light pseudoscalar meson ${\cal P}$ is obtained as 
\begin{eqnarray}\label{CFgluon}
\Big(\Pi^{\text{QCD(1)}}_{\mu \nu}\Big)_{\rho\sigma}(p,q) &=&  \frac{-i g_s}{96} \int \frac{d^4 k}{(2\pi)^2} e^{-i k.x} \int_0^1 du  \langle {\cal P}(q) \vert \bar{q}(0)\Gamma^J G_{\lambda \tau}q(x)\vert 0\rangle \nonumber\\ &\times&\Bigg\{\text{Tr}\Big[\tilde{\Delta}^{\lambda \tau}_{Q}(x) \gamma_\mu S^{(\text{pert.})}_{Q^{\prime}}(x) \gamma_\nu  \Big]\Big(\Gamma_{J}\Big)_{\rho \sigma} +
\text{Tr}\Big[  \gamma_\nu \tilde{\Delta}^{\lambda \tau}_{Q}(x)  \gamma_\mu  \Gamma_J  \Big] \Big(S^{(\text{pert.})}_{Q^{\prime}}(x)\Big)_{\rho \sigma}\nonumber
\\
&+&
\text{Tr} \Big[\Gamma_J  \gamma_\nu  \tilde{S}^{(\text{pert.})}_{Q^{\prime}}(x)  \gamma_\mu \Big] \Big(\Delta^{\lambda \tau}_{Q}(x)\Big)_{\rho \sigma}
+
\Big( \Gamma_J \gamma_\nu \tilde{\Delta}^{\lambda \tau}_{Q}(x) \gamma_\mu S^{(\text{pert.})}_{Q^{\prime}}(x) \Big)_{\rho \sigma}  \nonumber\\
&+&
\Big( \Gamma_J  \gamma_\nu   \tilde{S}^{(\text{pert.})}_{Q^{\prime}}(x) \gamma_\mu \Delta^{\lambda \tau}_{Q}(x) \Big)_{\rho \sigma} 
+
\Big(  S^{(\text{pert.})}_{Q^{\prime}}(x)  \gamma_\nu \tilde{\Delta}^{\lambda \tau}_{Q}(x)  \gamma_\mu \Gamma_J  \Big)_{\rho \sigma} 
\nonumber\\
&-&
\Big(S^{(\text{pert.})}_{Q^{\prime}}(x) \gamma_\nu \tilde{\Gamma}_J \gamma_\mu \Delta^{\lambda \tau}_{Q}(x)  \Big)_{\rho \sigma} -\Big(  \Delta^{\lambda \tau}_{Q}(x)  \gamma_\nu   \tilde{S}^{(\text{pert.})}_{Q^{\prime}}(x) \gamma_\mu \Gamma_J  \Big)_{\rho \sigma} \nonumber \\
&-&
\Big(\Delta^{\lambda \tau}_{Q}(x)   \gamma_\nu \tilde{\Gamma}_J \gamma_\mu S^{(\text{pert.})}_{Q^{\prime}}(x)  \Big)_{\rho \sigma} \Bigg\},
\end{eqnarray}
where the superscript $(1)$ indicates one gluon exchange. By exchanging $Q$ and $Q'$ in the above equation one can simply find the contribution of one gluon exchange between the heavy quark $Q'$ and the light pseudoscalar meson ${\cal P}$. 

The general configurations appear in the calculation of the QCD side of the CF (\ref{CFpert}) and (\ref{CFgluon}) have the form  
\begin{eqnarray}\label{Config1}
T_{[~~,\mu,\mu\nu,...]}(p,q)&=& i \int d^4 x \int_{0}^{1} dv  \int {\cal D}\alpha e^{ip.x} \big(x^2 \big)^n  [e^{i (\alpha_{q} + v \alpha _g) q.x} \mathcal{G}(\alpha_{i}) , e^{iq.x} f(u)] \nonumber\\
&\times& [1 , x_{\mu} , x_{\mu}x_{\nu},...]  K_{n_1}(m_1\sqrt{-x^2})  K_{n_2}(m_2\sqrt{-x^2}).
\end{eqnarray} 
The expressions in the brackets on the RHS correspond to different configurations which might arise in the calculation and 
\begin{equation}
\int \mathcal{D}\alpha =\int_{0}^{1}d\alpha _{q}\int_{0}^{1}d\alpha _{\bar{q}%
}\int_{0}^{1}d\alpha _{g}\delta (1-\alpha _{q}-\alpha _{\bar{q}}-\alpha
_{g}).
\end{equation}
On the LHS, the blank subscript indicates no $x_{\mu}$ in the corresponding configuration.
There are several representations for the modified Bessel function of the second kind and we use the cosine representation as
\begin{equation}\label{CosineRep}
K_n(m_Q\sqrt{-x^2})=\frac{\Gamma(n+ 1/2)~2^n}{\sqrt{\pi}m_Q^n}\int_0^\infty dt~\cos(m_Qt)\frac{(\sqrt{-x^2})^n}{(t^2-x^2)^{n+1/2}}.
\end{equation}
It is shown that choosing this representation increases the radius of convergence of the Borel transformed CF \cite{Azizi:2018duk}.
To perform the Fourier integral over $x$, we write the $x$ configurations in the exponential representation as 
\begin{eqnarray}\label{trick1}
(x^2)^n &=& (-1)^n \frac{\partial^n}{\partial \beta^n}\big(e^{- \beta x^2}\big)\arrowvert_{\beta = 0}, \nonumber \\
x_{\mu} e^{i P.x} &=& (-i) \frac{\partial}{\partial P^{\mu}} e^{i P.x}.
\end{eqnarray}

\subsection*{\textit{Borel Transformation and Continuum Subtraction}}
Performing the Fourier transformation, gives us the version of CF that has to be Borel transformed to  suppress  the divergences arise due to the dispersion integral. Having two independent momenta $(p+q)$ and $p$, we use the double Borel transformation with respect to the square of these momenta as
\begin{equation} \label{Borel1}
{\cal B}_{p_1}(M_{1}^{2}){\cal B}_{p_2}(M_{2}^{2})e^{b (p + u q)^2}=M^2 \delta(b+\frac{1}{M^2})\delta(u_0 - u) e^{\frac{-q^2}{M_{1}^{2}+M_{2}^{2}}},
\end{equation} 
where  $u_0 = M_{1}^{2}/(M_{1}^{2}+M_{2}^{2})$.
To be specific, we select the following configuration:
\begin{eqnarray}\label{Z1}
T_{\mu\nu}(p,q) &=& i \int d^4 x \int_{0}^{1} dv  \int {\cal D}\alpha e^{i[p+ (\alpha_{q} + v \alpha _g)q].x} \mathcal{G}(\alpha_{i}) \big(x^2 \big)^n  \nonumber\\
&\times& x_\mu x_\nu  K_{n_1}(m_{Q_1}\sqrt{-x^2})  K_{n_2}(m_{Q_2}\sqrt{-x^2}),
\end{eqnarray}
where after Fourier and Borel transformation leads to
\begin{eqnarray}\label{STR4}
T_{\mu\nu}(M^2) &=& \frac{i  \pi^2 2^{4-n_1-n_2} e^{\frac{-q^2}{M_1^2+M_2^2}}}{M^2 m_{Q_1}^{2n_1} m_{Q_2}^{2n_2}}\int  \mathcal{D}\alpha  \int_{0}^{1} dv \int_{0}^{1} dz \frac{\partial^n }{\partial \beta^n} e^{-\frac{m_{Q_1}^2 \bar{z} + m_{Q_2}^2 z}{z \bar{z}(M^2 - 4\beta)}} z^{n_1-1}\bar{z}^{n_2-1}  \nonumber\\
&\times & (M^2 - 4\beta)^{n_1+n_2-1} \delta[u_0 - (\alpha_{q} + v \alpha_{g})]  \Big[ p_\mu p_\nu + (v \alpha_{g} +\alpha_{q})(p_\mu q_\nu +q_\mu p_\nu ) \nonumber \\ 
&&+ (v \alpha_{g} +\alpha_{q})^2 q_\mu q_\nu + \frac{M^2}{2}g_{\mu\nu} \Big].
\end{eqnarray}
In Ref. \cite{Azizi:2018duk} one can find the details of calculations for hadrons containing different numbers of heavy quarks (zero to five).

The dispersion integral that matches the physical and QCD sides of the CF contains contributions from both the ground state and also the excited and continuum ones which is calculated in Eq. (\ref{CFPhys}) as a double dispersion integral over  $\rho^{\text{Phys.}}(s_1,s_2)$. To suppress the latter, one has to perform a proper version of subtraction which enhances the contribution of the ground state as well. 
There are several versions of subtraction.
Using quark-hadron duality one can approximate $\rho^{\text{Phys.}}(s_1,s_2)$, the physical spectral density, with its theoretical counterpart, $\rho^{\text{QCD}}(s_1,s_2)$. For more details see also  Refs. \cite{Li:2020rcg,Khodjamirian:2020mlb,Belyaev:1994zk, Aliev:2010yx, Aliev:2011ufa, Ball:1994}.

 To proceed, for the generic factor $(M^2)^N e^{-m^2/M^2}$, the  replacement \cite{Agaev:2016srl}
\begin{equation}
	\left( M^{2}\right) ^{N}e^{-m^{2}/M^{2}} \to
	\frac{1}{\Gamma (N)}\int_{m^{2}}^{s_0}dse^{-s/M^{2}}\left( s-m^{2}\right)
	^{N-1},
\end{equation}
is used  for $N>0$ and it is left unchanged for $N<0$, where the energy threshold for higher states and continuum is considered as $\sqrt{s_0}$.
Then we restrict the boundaries of $z$ integral in a way to more suppress the unwanted excited states and continuum. 
Solving the equation
\begin{eqnarray}\label{Sub1}
e^{-\frac{m_{Q_1}^2 \bar{z} + m_{Q_2}^2 z}{M^2 z \bar{z}}} = e^{-s_0/M^2},
\end{eqnarray}
for $z$ gives us the proper boundaries as follows:
\begin{eqnarray}\label{zlimits}
z_{\text{max}(\text{min})}=\frac{1}{2s_0}\Big[(s_0+m_{Q_1}^2-m_{Q_2}^2)+(-)\sqrt{(s0+m_{Q_1}^2-m_{Q_2}^2)^2-4m_{Q_1}^2s_0}\Big].
\end{eqnarray}
Replacing them in the $z$ integral as
\begin{eqnarray}\label{zsubtraction}
\int_{0}^{1}dz \rightarrow \int_{z_{\text{min}}}^{z_{\text{max}}}dz,
\end{eqnarray}
enhances the ground state and suppresses the higher states and continuum as much as possible.

Finally, selecting the coefficient of the corresponding structure $q_{\mu}q_{\nu}\rlap/p\rlap/q \gamma_5$, gives us the invariant function to be used to find the corresponding strong coupling constants.
The Borel transformed subtracted version of the invariant function for the vertex $B^*_1B^*_2{\cal P}$ can be written as 
\begin{eqnarray}\label{QCDfbs}
\Pi_{B^*_1B^*_2{\cal P}}(M^2,s_0) = T_{B^*_1B^*_2{\cal P}}(M^2,s_0)+T_{B^*_1B^*_2{\cal P}}^{GG}(M^2,s_0),
\end{eqnarray}
where the above functions  for the vertex $\Xi^*_{bb}\Xi^*_{bb}\pi^0$, as an example,  are given as
\begin{eqnarray}
T_{\Xi^*_{bb}\Xi^*_{bb}\pi^0}(M^2,s_0)&=&\frac{ e^{-\frac{q^2}{M_1^2+M_2^2}}}{9\sqrt{2} \pi ^2}  \Bigg\{6 u_0 f_{\pi } m_b m_{\pi }^2 i_1({\cal V}^\parallel(\alpha_i),1) \Big[\zeta
_{-1,0}(m_b)-\zeta _{0,0}(m_b)\Big]\nonumber\\
&+&6 u_0 f_{\pi} m_b m_{\pi}^2 i_1({\cal V}^\perp(\alpha_i),1) \Big[\zeta _{-1,0}(m_b)-\zeta
_{0,0}(m_b)\Big] \nonumber\\ &+& \mu_\pi \Bigg(-3 \Big[i_2(\alpha_g {\cal T}(\alpha_i),v)-2 i_2(\alpha_g {\cal T}(\alpha_i),v^2)+i_2(\alpha_q
{\cal T}(\alpha_i),1) \nonumber\\ &-&2 i_2(\alpha_q {\cal T}(\alpha_i),v)\Big] \Big[\tilde{\zeta}^{(1)} _{0,0}(m_b;s_0,M^2)-\tilde{\zeta}^{(1)} _{1,0}(m_b;s_0,M^2)\Big]\\
&+&(-1+\rho_\pi^2)
u_0^2 \phi_\sigma(u_0) \Big[-\tilde{\zeta}^{(1)}_{1,0}(m_b;s_0,M^2)+\tilde{\zeta}^{(1)}_{2,0}(m_b;s_0,M^2)\Big]\Bigg)\Bigg\}\nonumber,
\end{eqnarray}
and
\begin{eqnarray}
T_{\Xi^*_{bb}\Xi^*_{bb}\pi^0}^{GG}(M^2,s_0)&=&g_s^2\langle GG \rangle\frac{ e^{-\frac{q^2}{M_1^2+M_2^2}}}{324 \sqrt{2}M^6 \pi ^2} \Bigg\{6 u_0 f_{\pi} m_b^3 m_{\pi }^2 i_1({\cal V}^\parallel(\alpha_i),1)
\Big[\zeta _{-2,0}(m_b)+\zeta _{-1,-1}(m_b)\Big]\nonumber\\
&+&6 u_0 f_{\pi } m_b^3 m_{\pi}^2 i_1({\cal V}^\perp(\alpha_i),1)
\Big[\zeta _{-2,0}(m_b)+\zeta _{-1,-1}(m_b)\Big] \nonumber \\ &+& \mu_{\pi} \Bigg(-3 \Big[i_2(\alpha_g {\cal T}(\alpha_i),v)-2i_2(\alpha_g {\cal T}(\alpha_i),v^2)+i_2(\alpha_q {\cal T}(\alpha_i),1)\nonumber\\
&-&2 i_2(\alpha_q {\cal T}(\alpha_i),v)\Big] \Big\{m_b^2
\Big[\tilde{\zeta}^{(1)}_{-1,0}(m_b;s_0,M^2)+\tilde{\zeta}^{(1)}_{0,-1}(m_b;s_0,M^2)\Big] \nonumber\\ &+&\tilde{\zeta}^{(2)}_{0,0}(m_b;s_0,M^2)\Big\} -2 (-1+\rho_\pi^2)
u_0^2 \phi_\sigma(u_0) \Big[m_b^2 \tilde{\zeta}^{(1)}_{-1,0}(m_b;s_0,M^2) \nonumber\\
&+&m_b^2 \tilde{\zeta}^{(1)}_{0,-1}(m_b;s_0,M^2)+2 \tilde{\zeta}^{(2)}_{0,0}(m_b;s_0,M^2)  -m_b^2
\tilde{\zeta}^{(1)}_{0,0}(m_b;s_0,M^2) \nonumber\\ 
&-&m_b^2 \tilde{\zeta}^{(1)}_{1,-1}(m_b;s_0,M^2)-2 \tilde{\zeta}^{(2)}_{1,0}(m_b;s_0,M^2)\Big]\Bigg)\Bigg\}.
\end{eqnarray} 
Here the function $\zeta_{m,n}(m_1,m_2)$ is defined as
\begin{eqnarray}
\zeta_{m,n}(m_1,m_2) = \int_{z_{\text{min}}}^{z_{\text{max}}}dz z^m \bar{z}^n e^{-\frac{m_1^2}{M^2 z}-\frac{m_2^2}{M^2 \bar{z}}}~,
\end{eqnarray}
where $m$ and $n$ are integers and $\zeta_{m,n}(m_1,m_1)=\zeta_{m,n}(m_1)$.
Also $\tilde{\zeta}^{(N)}_{m,n}(m_1,m_1;s_0,M^2)$ is defined as follows:
\begin{eqnarray}
	\tilde{\zeta}^{(N)}_{m,n}(m_1,m_2;s_0,M^2) =\frac{1}{\Gamma(N)} \int_{z_{\text{min}}}^{z_{\text{max}}}dz\int_{\frac{m_1^2}{z}+\frac{m_2^2}{\bar{z}}}^{s_0} ds e^{-s/M^2}z^m \bar{z}^n\Big(s-\frac{m_1^2}{z}-\frac{m_2^2}{\bar{z}}\Big)^{N-1}~,
\end{eqnarray}
where $\tilde{\zeta}^{(N)}_{m,n}(m_1,m_1;s_0,M^2)=\tilde{\zeta}^{(N)}_{m,n}(m_1;s_0,M^2)$.
The functions $i_1$ and $i_2$ are also defined as
\begin{eqnarray}
i_1(\phi(\alpha_i),f(v)) = \int {\cal D}\alpha_i \int_{0}^{1}dv \phi(\alpha_{q},\alpha_{\bar{q}},\alpha_{g}) f(v) \theta(k-u_0)~,
\end{eqnarray}
\begin{eqnarray}
i_2(\phi(\alpha_i),f(v)) = \int {\cal D}\alpha_i \int_{0}^{1}dv \phi(\alpha_{q},\alpha_{\bar{q}},\alpha_{g}) f(v) \delta(k-u_0)~,
\end{eqnarray}
where $k=\alpha_{q} - v\alpha_{g}$.

Finally, matching both the physical  and QCD  sides of the CF  leads to the sum rules for the strong coupling constant as
\begin{eqnarray}\label{SumRule}
g_{B^*_1B^*_2{\cal P}} = \frac{3 m_{1}^2}{2} \dfrac{e^{m_{1}^2/M_1^2} e^{m_{2}^2/M_2^2}}{\lambda_{B^*_1} \lambda_{B^*_2}} \Pi_{B^*_1B^*_2{\cal P}}(M^2,s_0).
\end{eqnarray}

\section{Numerical Results}\label{NA}

In this section, we present the numerical results of the sum rules for the strong coupling
constants of light pseudoscalar meson with doubly heavy spin–$ 3/2 $ baryons.
For numerical evaluations, two sets of parameters are being employed as follows: 
One which corresponds to light pseudoscalar mesons that are  their masses and decay constants as well as their non-perturbative parameters
appearing in their DAs of different twists, which are listed in Tables (\ref{tabmass}) and (\ref{tabDAs}) respectively. The other is
the masses and residues of the doubly heavy  spin-3/2
baryons and are listed in Table (\ref{tabBaryon}).
\begin{table}[t]
	\renewcommand{\arraystretch}{1.3}
	\addtolength{\arraycolsep}{1pt}
	$$
	\rowcolors{2}{olive!10}{white}
	\begin{tabular}{|c|c|c|c|}
	\rowcolor{olive!30}
	\hline \hline
	\mbox{Parameters}         &  \mbox{Values  }  
	\\
	\hline\hline
	$ m_s $    & $  93^{+11}_{-5}~\mbox{MeV} $      \\
	$ m_{c} $    &  $ 1.27^{+0.02}_{-0.02}~\mbox{GeV} $      \\
	$ m_b $    & $  4.18^{+0.03}_{-0.02}~\mbox{GeV} $      \\
	$ m_{\pi^0} $    &   $134.9768\pm0.0005 ~\mbox{MeV}$  \\
	$ m_{\pi^\pm} $    &   $139.57039\pm0.00018 ~\mbox{MeV}$  \\
	$ m_{\eta } $    &   $957.862\pm0.017 ~\mbox{MeV}$  \\
	$ m_{K^0} $    &   $497.611\pm0.013~\mbox{MeV} $  \\
	$ m_{K^{\pm}}    $  &   $493.677\pm0.016~\mbox{MeV}$    \\
	$  f_\pi $    & $  131~ \mbox{MeV} $     \\
	$  f_\eta $    & $  130~\mbox{MeV}  $     \\
	$  f_{K} $    & $  160~ \mbox{MeV} $     \\
	\hline \hline
	\end{tabular}
	$$
	\caption{The meson masses and leptonic decay constants along with the  quark masses \cite{Tanabashi:2018oca}.}  \label{tabmass} 
	\renewcommand{\arraystretch}{1}
	\addtolength{\arraycolsep}{-1.0pt}
\end{table}

\begin{table}[t]
	\renewcommand{\arraystretch}{1.3}
	\addtolength{\arraycolsep}{1pt}
	$$
	\rowcolors{2}{olive!10}{white}
	\begin{tabular}{|c|c|c|c|c|c|}
	\rowcolor{olive!30}
	\hline \hline
	\text{meson}  &  $a_2$ & $\eta_3$ & $w_3$ & $\eta_4$ & $w_4$  
	\\
	\hline\hline
	$\pi$  & 0.44 & $0.015$ & -3 & 10 & 0.2     \\
	
	$ $K$ $ & 0.16  & $0.015$ & -3 & 0.6 & 0.2     \\
	$\eta $ & 0.2 & 0.013 & -3 & 0.5 & 0.2     \\
	\hline \hline
\end{tabular}
$$
\caption{Input parameters for DAs of twist 2, 3 and 4 at the renormalization scale $\mu=1\text{GeV}$ \cite{Ball:2005vx,Ball:2004ye}.}  \label{tabDAs} 
\renewcommand{\arraystretch}{1}
\addtolength{\arraycolsep}{-1.0pt}
\end{table}

\begin{table}[t]
	\renewcommand{\arraystretch}{1.3}
	\addtolength{\arraycolsep}{1pt}
	$$
	\rowcolors{2}{olive!10}{white}
	\begin{tabular}{|c|c|c|c|}
	\rowcolor{olive!30}
	\hline \hline
	\mbox{Baryon}         &  \mbox{Mass} $[\text{GeV}]$  & \mbox{Residue $[\text{GeV}^3]$}
	\\
	\hline\hline
	
	$ \Xi^*_{cc}  $    &  $ 3.69\pm0.16 $  & $0.12\pm0.01$ \\
	$ \Xi^*_{bc}  $    &  $ 7.25\pm0.20 $ & $0.15\pm0.01$  \\
	$ \Xi^*_{bb} $    &  $ 10.14\pm1.0 $  & $0.22\pm0.03$  \\
	$ \Omega^*_{cc}  $    &  $ 3.78\pm0.16 $  & $0.14\pm0.02$ \\
	$ \Omega^*_{bc}  $    &  $ 7.3\pm0.2 $  & $0.18\pm0.02$ \\
	$ \Omega^*_{bb} $    &  $ 10.5\pm0.2  $  & $0.25\pm0.03$ \\
	
	\hline \hline
\end{tabular}
$$
\caption{The  masses and residues of spin-$ 3/2 $ doubly heavy baryons  \cite{Aliev:2012iv}.}  \label{tabBaryon} 
\renewcommand{\arraystretch}{1}
\addtolength{\arraycolsep}{-1.0pt}
\end{table}

The first step in the numerical analysis is to fix the working intervals of the auxiliary parameters. These are  continuum threshold $ s_0 $ and Borel parameter $ M^2 $. To this end, we impose the conditions of: OPE convergence, pole dominance and relatively weak dependence of the results on  auxiliary parameters. The threshold $s_0$,   depends on the energy of the first excited state at each channel. We have not any experimental information on the excited states of the doubly heavy baryons, however, our analyses show that in the interval $ m_{B^*}+  0.3~ \text{GeV} \leq\sqrt{s_0}\leq  m_{B^*}+ 0.5 ~\text{GeV} $ the standard requirements are satisfied. To fix  the Borel parameter $ M^2 $ we follow the steps: for its upper limit we impose the condition of pole dominance and for determination of its lower limit we demand the OPE convergence condition.  These requirements lead to
the working regions $14$~GeV$ ^2  $ $\leq M^2$ $\leq 18$~GeV $^2 $, $8$~GeV$ ^2  $ $\leq M^2$ $\leq 11$~GeV $^2 $ and $4$~GeV$ ^2  $ $\leq M^2$ $\leq 6$~GeV $^2 $ for $ bb $, $ bc $ and $ cc $ channels, respectively.

In Fig. (\ref{fig:s0}), as an example, we show the dependence of the coupling constant $ g $ on the continuum threshold at some fixed values of the  Borel parameter for the vertices $ \Xi^*_{bb}\Xi^*_{bb} \pi^{\pm} $, $ \Omega^*_{bb}\Omega^*_{bb} \eta $, $ \Omega^*_{bb}\Xi^*_{bb} K^{\pm} $, and $ \Omega^*_{bb}\Xi^*_{bb} K^0 $. As is seen, the strong coupling constants depend very weakly on $ s_0 $.
\begin{figure}[h!]
	\includegraphics[width=1.0\textwidth]{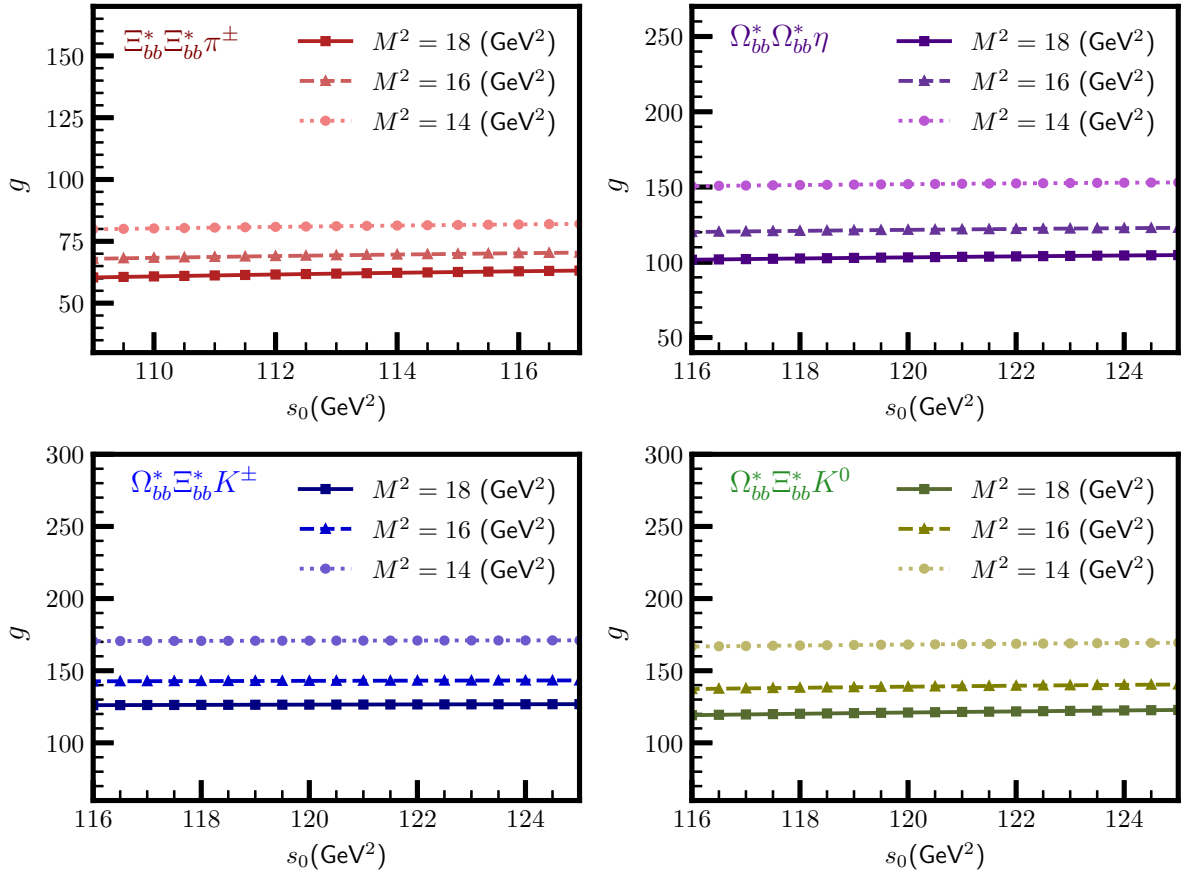}
	\caption{Dependence of the strong coupling constant $g$  on the continuum threshold at different values of the  Borel parameter $ M^2 = 14, 16 $ and $ 18 $~GeV$ ^2 $.}
	\label{fig:s0}
\end{figure} 
As another example, we show the dependence of $ g $ on the Borel parameter $ M^2 $ for the vertices $ \Xi^*_{bb}\Xi^*_{bb} \pi^{\pm} $, $ \Omega^*_{bb}\Omega^*_{bb} \eta $, $ \Omega^*_{bb}\Xi^*_{bb} K^{\pm} $, and $ \Omega^*_{bb}\Xi^*_{bb} K^0 $ at fixed values of $s_0$  in Fig (\ref{fig:msq}). From these plots, we see that the strong coupling constants show some dependence on the Borel parameter, which constitute the main parts of the uncertainties. These uncertainties, however, remain inside the limits allowed by the method.
\begin{figure}[h!]
	\includegraphics[width=1.0\textwidth]{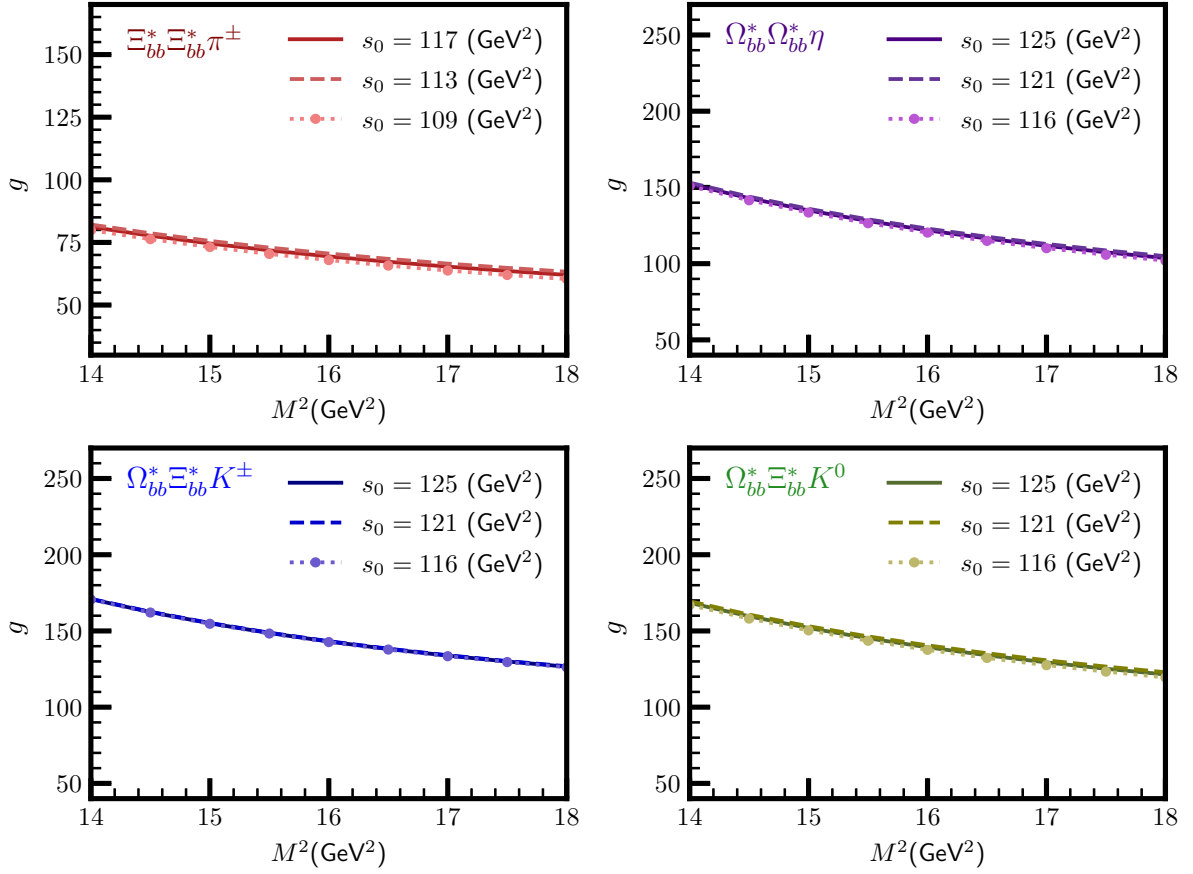}
	\caption{Dependence of the strong coupling constant $g$ on the Borel parameter at different values of the continuum threshhold $s_0$. The lines correspond to $ s_0  = 109$, $ 113 $ and $ 117 $~GeV$ ^2 $ for $ \Xi^*_{bb}\Xi^*_{bb} \pi^{\pm} $ channel and  $ s_0  = 116$, $ 121 $ and $ 125 $~GeV$ ^2 $ for $ \Omega^*_{bb}\Omega^*_{bb} \eta $, $ \Omega^*_{bb}\Xi^*_{bb} K^{\pm} $, and $ \Omega^*_{bb}\Xi^*_{bb} K^0 $ones.}
	\label{fig:msq}
\end{figure}

Taking into account all the input values as well as the working intervals for the auxiliary parameters, we extract the numerical values for the strong coupling constants from the light-cone sum rules (\ref{SumRule}). The numerical results of $g$ for all the considered vertices in $ cc $, $ bc $ and $ bb $  channels are presented in Table (\ref{tab:g}).
\begin{table}[h!]
	
	\begin{center}
		\rowcolors{2}{olive!10}{white}
		\begin{tabular}{ |c c c c |}
			\rowcolor{olive!30}
			\hline
			\hline
			\mbox{Vertex}  & $ M^2 $(GeV$ ^2 $) & $ s_0 $ (GeV$ ^2 $) & \mbox{strong coupling constant}\\
			\hline
			\hline
			\multicolumn{4}{|c|}{  \textcolor{olive}{Decays to $ \pi $}} \\
			\hline
			$ \Xi^*_{bb} \Xi^*_{bb} \pi^0  $&$ 14\leq M^2\leq18 $ &$ 109\leq s_0\leq117 $ &$  50.0^{\:7.9}_{\:7.8}  $\\
			$ \Xi^*_{bb} \Xi^*_{bb} \pi^\pm  $&$ 14\leq M^2\leq18 $ &$  109\leq s_0\leq117 $ &$  70.8^{\:11.4}_{\:10.3}  $\\
			\hline 
			$ \Xi^*_{bc}\Xi^*_{bc} \pi^0  $&$ 7\leq M^2\leq10 $ &$57\leq s_0\leq63$ &$13.4^{\:1.4}_{\:1.4}   $\\
			$ \Xi^*_{bc} \Xi^*_{bc} \pi^\pm  $&$ 7\leq M^2\leq10 $ &$ 57\leq s_0\leq63$ &$18.9^{\:2.0}_{\:2.0}  $ \\
			\hline
			$ \Xi^*_{cc} \Xi^*_{cc} \pi^0  $ & $ 3\leq M^2\leq6 $ &$ 16\leq s_0\leq19 $ & $ 3.8^{\:0.3}_{\:0.4} $ \\
			$ \Xi^*_{cc} \Xi^*_{cc} \pi^\pm  $ & $ 3\leq M^2\leq6 $ &$ 16\leq s_0\leq19$ & $5.5 ^{\:0.5}_{\:0.6} $\\
			\hline
			\hline
			\multicolumn{4}{|c|}{  \textcolor{olive}{Decays to $\eta$}} \\
			\hline
			$ \Omega^*_{bb}  \Omega^*_{bb} \eta  $&$ 14\leq M^2\leq18 $ &$  116\leq s_0\leq125$ &$ 125.7^{\:22.3}_{\:23.7}  $\\
			$ \Omega^*_{bc}  \Omega^*_{bc} \eta $& $ 7\leq M^2\leq10 $ &$57\leq s_0\leq64 $ &$ 18.4^{\:2.1}_{\:2.0} $\\
			$ \Omega^*_{cc} \Omega^*_{cc} \eta $&$ 3\leq M^2\leq6 $ &$ 16\leq s_0\leq20 $ &$ 5.9^{\:0.5}_{\:0.6} $\\
			\hline
			\hline
			\multicolumn{4}{|c|}{ \textcolor{olive}{Decays to $K$}} \\
			\hline
			$ \Omega^*_{bb} \Xi^*_{bb} K^0  $&$ 14\leq M^2\leq18 $ &$  116\leq s_0\leq125$ &$142.9^{\:26.6}_{\:23.4}  $\\
			$ \Omega^*_{bb} \Xi^*_{bb} K^\pm  $& $ 14\leq M^2\leq18 $ &$ 116\leq s_0\leq125$ &$ 142.3^{\:26.4}_{\:23.2} $\\
			\hline
			$ \Omega^*_{bc} \Xi^*_{bc} K^0  $&$ 7\leq M^2\leq10 $ &$57\leq s_0\leq64$ &$27.5^{\:3.0}_{\:2.9}    $\\
			$ \Omega^*_{bc} \Xi^*_{bc} K^\pm  $& $ 7\leq M^2\leq10 $ &$57\leq s_0\leq64$ &$  27.3^{\:3.0}_{\:2.9} $\\
			\hline
			$ \Omega^*_{cc} \Xi^*_{cc} K^0  $&$ 3\leq M^2\leq6 $ &$ 16\leq s_0\leq20 $ &$  8.7^{\:0.8}_{\:0.9} $\\
			$ \Omega^*_{cc} \Xi^*_{cc} K^\pm  $& $ 3\leq M^2\leq6 $ &$ 16\leq s_0\leq20 $ &$  8.6^{\:0.7}_{\:0.9} $\\
			\hline
		\end{tabular}
	\end{center}
	\caption{Working regions of the Borel parameter $ M^2 $ and   continuum threshold $ s_0 $  as well as the numerical values for different strong coupling constants extracted from the analyses.} \label{tab:g} 
\end{table}
There are three sources of errors in the results. The first is from the input parameters we have used in the calculations. The second, the intrinsic error, is due to the employment of the quark-hadron duality assumption and finally, the third one, is the error because of the variations with respect to the auxiliary parameters in their working intervals. The results contain  errors overall up to $20 \%$.

\newpage
\section{Summary and Conclusions}\label{SC}
In the present study, we investigated the strong coupling constants of the doubly heavy  spin-$ 3/2 $ baryons, $\Omega^*_{QQ'}  $ and  $\Xi^*_{QQ'}  $, with the  light pseudoscalar mesons, $ \pi$,  $K$ and  $ \eta$, by applying the LCSR formalism. We used the interpolating currents of the doubly heavy baryons and DAs of the mesons to  construct the sum rules for the strong coupling constants. By fixing the auxiliary parameters, entered  the calculations through the Borel transformation as well as continuum subtraction, we extracted the numerical results for different possible  vertices considering the quark contents of the participating particles and other considerations. 

The obtained results indicate that the heavier the doubly heavy baryons, the larger their strong couplings to the same pseudoscalar mesons.
The significant difference between the c- and b-baryons can be explained using the heavy-quark spin symmetry that tells us the partial strong decay widths of heavy quark hadronic systems are independent of the heavy quark flavor and spin \cite{IsW}. Considering the Fermi golden rule and the fact that the amplitude is proportional to the strong coupling constant, defining the corresponding vertex, immediately implies that the heavier doubly-heavy baryon has the larger strong coupling constant.
Also comparing with the results presented in Ref. \cite{Rostami:2020euc}, one finds out that the values of the strong coupling constants among the doubly heavy spin-3/2 baryons and light pseudoscalar mesons are mostly larger than that of the spin-1/2 baryons with the same mesons taken into account in Ref. \cite{Rostami:2020euc}.

Our results may help experimental groups in analyses of the data including the hadrons taken into account in the present study. The results may also be used in the construction of the strong potentials among the doubly heavy baryons and pseudoscalar mesons.

\section*{ACKNOWLEDGEMENTS}
K. Azizi and S. Rostami are thankful to Iran Science Elites Federation (Saramadan)
for the financial support provided under the grant number ISEF/M/99171.

\end{document}